\documentclass[12pt]{iopart}

\usepackage{graphicx}
\begin{document}

\def\no{\noindent}
\def\non{\nonumber}
\def\bear{\begin{eqnarray}}
\def\ear{\end{eqnarray}}
\def\half{{1\over 2}}
\def\fourth{{1\over 4}}
\def\mn{\mu\nu}
\def\Tintm{{\dps\int_{0}^{\infty}}{dT\over T}\,e^{-m^2T}}
\def\TintmD{{\dps\int_{0}^{\infty}}{dT\over T}\,e^{-m^2T}
    {(4\pi T)}^{-{D\over 2}}}
\def\dps{\displaystyle}
\def\exmn{\Bigl(\mu \leftrightarrow \nu \Bigr)}
\def\pa{\partial}

\setlength{\topmargin}{-1.2cm}

\pagestyle{empty}

\begin{center}
{\huge \bf  Photon-graviton mixing in an electromagnetic field}

\bigskip
\bigskip
\bigskip

\no
{\bf F. Bastianelli$^1$, U. Nucamendi$^2$, \underline{C. Schubert}$^2$, V.M. Villanueva$^2$}

\end{center}

\medskip

{\it \noindent $^1$
Dipartimento di Fisica, Universit\`a di Bologna and INFN, Sezione di Bologna,
Via Irnerio 46, I-40126 Bologna, Italy}

\medskip

{\it \noindent $^2$
Instituto de F\'{\i}sica y Matem\'aticas,
Universidad Michoacana de San Nicol\'as de Hidalgo,
Edificio C-3, Apdo. Postal 2-82,
C.P. 58040, Morelia, Michoac\'an, M\'exico}

\bigskip 
\bigskip
\bigskip

\no
Talk given by C. S. at {\it 8th Workshop on Quantum Field Theory under the Influence of External Conditions} (QFEXT07), Leipzig, 17-21 Sep 2007.

\bigskip
\bigskip
\bigskip
\bigskip

\no
{\bf Abstract:} Einstein-Maxwell theory implies the mixing of photons with
gravitons in an external electromagnetic field. This process and
its possible observable consequences have been
studied at tree level for many years. We use the worldline formalism for
obtaining an exact integral representation for the one-loop corrections
to this amplitude due to scalars and fermions. We study the structure of
this amplitude, and obtain exact expressions for various limiting cases.



\vfill\eject\

\pagestyle{plain}

\setcounter{page}{1}


\section{Photon-graviton mixing at tree level and one loop}

As has been realized already in the sixties 
\cite{gertsenshtein}
Einstein-Maxwell theory in a constant electromagnetic
field contains a tree level vertex for photon-graviton conversion,


\bear
\kappa h_{\mn}F^{\mu\alpha}f^{\nu}_{\,\,\,\alpha} 
- \fourth\kappa h^{\mu}_{\mu}F^{\alpha\beta} f_{\alpha\beta}.
\label{vertex}
\ear
Here $h_{\mu\nu}$  denotes the graviton, $f_{\mu\nu}$ 
 the photon, and $F^{\mu\nu}$ 
the external field.
$\kappa$  is the gravitational coupling constant.
This vertex implies the possibility of photon-graviton oscillations 
\cite{gertsenshtein,zelnovbook,rafsto,magueijo,chen,cilhar,defuza}
which are analogous to the better-known 
photon--axion oscillations in a field \cite{primakoff}.  
In the presence of an external field, the
true eigenstates of propagation can be obtained
by solving the following system of dispersion relations
\cite{rafsto}

\bear
\Biggl(
\matrix {\eta^{\alpha\beta}k^2-k^{\alpha}k^{\beta}&
{i\over 2}\kappa C^{\kappa\lambda,\alpha} 
\cr
{i\over 2}\kappa C^{\mn,\beta}&
{k^2\over 4}(\eta^{\mu\kappa}\eta^{\nu\lambda}
\!+\!
\eta^{\mu\lambda}\eta^{\nu\kappa}
\!-\!2\eta^{\mn}\eta^{\kappa\lambda} + \! {\scriptstyle \ldots}) 
\! \cr
}
\Biggr) 
\Biggl(
\matrix {a_{\beta}(k) \cr
h_{\kappa\lambda}(k) \cr
}
\Biggr)  
&& = 0 \, .
\label{disptree}
\ear
Here $C^{\mn,\alpha}$ denotes the Fourier transform of the vertex (\ref{vertex}),

\bear
C^{\mn,\alpha} &=&
\bigl(F\cdot k\bigr)^{\alpha}\eta^{\mn}
+F^{\mu\alpha}k^{\nu}
+F^{\nu\alpha}k^{\mu}
-\bigl(F\cdot k\bigr)^{\mu}\eta^{\nu\alpha}
-\bigl(F\cdot k\bigr)^{\nu}\eta^{\mu\alpha} \, 
.
\label{defC}
\ear
For small deviations from the vacuum dispersion relations $k^2=0$ this second order
equation can be linearized. An efficient formalism for solving the system (\ref{defC}) under this
condition was developed in \cite{rafsto}.

Taking one--loop corrections into account, the dispersion relation matrix gets modified
in the following way \cite{pg1,pg2},

\bear
\fl\biggl(
\matrix {\eta^{\alpha\beta}k^2-k^{\alpha}k^{\beta}- 
\bar\Pi^{\alpha,\beta}& 
{i\over 2}\kappa C^{\kappa\lambda,\alpha} -
 \bar\Pi^{\kappa\lambda,\alpha} 
\cr
{i\over 2}\kappa C^{\mn,\beta}-\bar\Pi^{\mn,\beta}&
{k^2\over 4}(\eta^{\mu\kappa}\eta^{\nu\lambda}
\!+\!
\eta^{\mu\lambda}\eta^{\nu\kappa}
\!-\!2\eta^{\mn}\eta^{\kappa\lambda} + \! {\scriptstyle \ldots}) 
\!-\!\bar\Pi^{\mn,\kappa\lambda} \cr
}
\biggr) 
\Biggl(
\matrix {a_{\beta}(k) \cr
h_{\kappa\lambda}(k) \cr
}
\Biggr) =0\, . 
\non\\
\label{disploop}
\ear
Here $\bar\Pi^{\alpha,\beta}$, $\bar\Pi^{\mn,\beta}$, and
$\bar\Pi^{\mn,\kappa\lambda}$ denote the one--loop photon--photon,
graviton--photon, and graviton--graviton vacuum polarization tensors
in a constant field. 
These quantities in principle have to be calculated with all possible
loop particles.
Eqs. (\ref{disploop}) generalize the QED dispersion relation 


\bear
\Bigl(\eta^{\alpha\beta}k^2-k^{\alpha}k^{\beta}- 
\bar\Pi^{\alpha,\beta}\Bigr)
\, a_{\beta}(k) = 0 \, .
\label{QEDdisp}
\ear
This case is well-known and has been studied by many authors 
(see, e.g., \cite{adler,batsha, tsaerb}).
It leads to a complicated dependence of the phase velocity on
polarization, field strength and frequency (see \cite{ditgiebook}
for a detailed discussion).

\begin{figure}[h]
\centering
\includegraphics{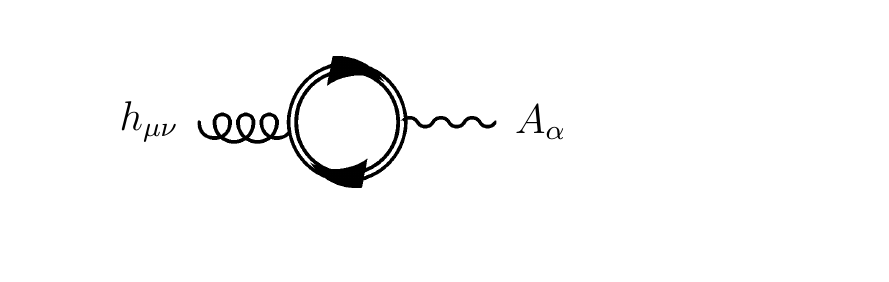}
\vspace{-14pt}
\caption{One-loop photon-graviton amplitude in a constant field. The double line represents
the propagator of a charged scalar or spin $\half$ particle in a constant field.}
\label{fig1}
\end{figure}

In this talk, we report on our recently concluded calculation \cite{pg1,pg2} of the 
photon-graviton polarization tensor in a constant electromagnetic field
 $\bar \Pi^{\mu\nu,\alpha}$, with a charged scalar or spin $\half$ particle in the loop.
As a Feynman diagram, this amplitude is represented by fig. \ref{fig1}.


In \cite{pg1} the worldline formalism 
was used to obtain compact parameter integral representations
for this amplitude. The numerical and structural analysis has been concluded
only recently \cite{pg2}.
Since this formalism is presently still somewhat novel, particularly
in applications to gravity, we will start with shortly reviewing its basics from a user's point
of view.

\vspace{-8pt}

\section{Worldline formalism in a constant electromagnetic field}

The worldline formalism goes back to Feynman's representation of
scalar \cite{feyn50} and spinor \cite{feyn51} QED in terms of
relativistic particle path integrals. Let us write down Feynman's
integral for the simplest possible case, the one-loop effective
action in scalar QED:

\vspace{-10pt}

\bear
\Gamma(A) &=&
\int d^4x\, {\cal L}(A) =
\int_0^{\infty}{dT\over T}\,{\rm e}^{-m^2T}
{\displaystyle \int_{x(T)=x(0)}}{\cal D}x(\tau)
\, e^{-S[x(\tau)]} \, .
\label{Gammascal}
\ear
Here $m$ and $T$ denote the mass and proper time of the loop scalar.
The worldline path integral is to be calculated over loops in spacetime with
fixed periodicity $T$, and a worldline action given by


\bear
S[x(\tau)] \, & =& \int_0^T d\tau
\biggl[
\fourth \,\dot x^2 + 
 ie  \dot x^{\mu}A_{\mu}(x(\tau)) 
 \biggr] \, .
\label{Swlscal}
\ear
In the fermion QED case, a number of different ways have been found to
implement the spin in the worldline path integral. Feynman's original
formulas \cite{feyn51} are based on a spin factor involving Dirac matrices,
but for the purpose of analytic calculation it is usually preferable to implement
spin by the following additional Grassmann path integral
\cite{fradkin66},

\bear
\int {\cal D}\psi(\tau)
\,
\exp 
\Biggl\lbrack
-\int_0^Td\tau
\Biggl(
\half \psi\cdot \dot\psi -ie \psi^{\mu}F_{\mn}\psi^{\nu}
\Biggr)
\Biggr\rbrack \, .
\label{grassmannpi}
\ear
Here the functions $\psi^{\mu}$ are anticommuting and
antiperiodic,


\bear
\psi(\tau_1)\psi(\tau_2) &=& - \psi(\tau_2)\psi(\tau_1), \qquad
\psi(T) = - \psi(0) \, .
\label{antianti}
\ear
See, e.g.,  chapter 3 of \cite{review} for a derivation of the path integral representations
(\ref{Gammascal}), (\ref{grassmannpi}) from quantum field theory.
During the last
fifteen years various efficient methods have been developed for the
evaluation of this type of path integral. We are concerned here with the
so-called ``string-inspired'' approach 
\cite{polbook,berkos,strassler} (see \cite{review} for a review) 
which aims at an analytic calculation of the worldline path integral
(see the contributions by G.V. Dunne and K. Klingm\"uller to
these proceedings for alternative approaches).
This is achieved by manipulating the path integral into gaussian form,
usually by a perturbative or higher derivative expansion, and then 
performing the gaussian integration formally using worldline correlators.
Those worldline Green's functions are, for the coordinate path integral,


\bear
\fl\qquad \langle x^{\mu}(\tau_1)x^{\nu}(\tau_2) \rangle
&=&
-G_B(\tau_1,\tau_2)\, \eta^{\mu\nu}, \quad
G_B(\tau_1,\tau_2) = \vert \tau_1 -\tau_2\vert -{\Bigl(\tau_1 -\tau_2\Bigr)^2\over T},\non\\
\label{GB}
\ear
and for the spin path integral


\bear
\langle \psi^{\mu}(\tau_1)\psi^{\nu}(\tau_2)\rangle
&=&
G_F(\tau_1,\tau_2)\, \eta^{\mu\nu}, \quad
G_F(\tau_1,\tau_2) = {\rm sign}(\tau_1 - \tau_2).
\label{GF}
\ear
We will often abbreviate $G_B(\tau_1,\tau_2)=:G_{B12}$ etc.
The coordinate Green's function is not unique, since it depends on the
zero mode fixing of the path integral \cite{review}. 
The choice of (\ref{GB}) corresponds to a definition of
the zero mode as the loop center-of-mass,

\bear
x_0^{\mu} &:=& {1\over T}\int_0^Td\tau \, x^{\mu}(\tau) \, .
\label{zeromode}
\ear
To obtain, say, the one-loop $N$ photon amplitude, 
one expands the Maxwell field in $N$ plane waves with
given polarization vectors $\varepsilon_i^{\mu}$,

\bear
A^{\mu}(x(\tau)) &=& \sum_{i=1}^N \,\varepsilon_i^{\mu}\e^{ik_i\cdot x(\tau)} \, .
\label{Aexpand}
\ear
This leads to each photon being represented by a 
photon vertex operator, 


\bear
V_{\rm scal}^A [k,\varepsilon]
&=&\int_0^T d\tau\,  \varepsilon\cdot \dot x(\tau)\,\e^{ik_i\cdot x(\tau)} 
\hspace{103pt} (Scalar\,\, QED) \, ,
\nonumber\\
V_{\rm spin}^A [k,\varepsilon]
&=&\int_0^T d\tau\,  
\Bigl[\varepsilon\cdot \dot x(\tau) +2i\varepsilon\cdot\psi k\cdot\psi\Bigr] 
\,\e^{ik_i\cdot x(\tau)}
\qquad (Spinor\,\, QED) \, .
 \nonumber\\
\label{photvertop}
\ear


\no
After expanding the interaction exponential to $N$th order, the path integrals are 
gaussian and can be evaluated by the correlators (\ref{GB}), (\ref{GF}). This leaves one with
the global proper time integral, and one parameter integral for each photon leg.

It has emerged that this formalism is particularly well-suited to the calculation of
QED amplitudes in a constant background field \cite{Fext,vv}.
The reason is that, once one has obtained the parameter integral representation
for a given amplitude for the vacuum case, one can construct the corresponding
integrals in the presence of a background field with constant field strength
tensor $F_{\mu\nu}$ by the follwing simple
substitutions \cite{vv,review}:


\begin{itemize}

\item
\no
Change the worldline Green's functions:

\bear
\fl \qquad G_B(\tau_1,\tau_2)&\to&{\cal G}_B(\tau_1,\tau_2) =
{1\over 2{(eF)}^2}\Biggl({eF\over{{\rm sin}(eFT)}}
{\rm e}^{-ieFT\dot G_{B12}}
\!+\! ieF\dot G_{B12}\! -\!{1\over T}\Biggr)\, , \non\\
&&\non\\
\fl \qquad G_F(\tau_1,\tau_2) &\to& {\cal G}_F(\tau_1,\tau_2) =
G_{F12}\,
{{\rm e}^{-ieFT\dot G_{B12}}\over {\rm cos}(eFT)} \, .
\non\\
\label{changegreen}
\ear
(A `dot' on a Green's function denotes a derivative with respect to the first variable.)

\item
\no
Change the free path integral determinants:


\bear
(4\pi T)^{-{D\over 2}} &\to&  (4\pi T)^{-{D\over 2}} \,
{\rm det}^{-\half}\Biggl\lbrack{\sin (eFT)\over eFT}\Biggr\rbrack
\qquad  (Scalar \,\, QED) \, ,
\nonumber\\
&&\nonumber\\
(4\pi T)^{-{D\over 2}} &\to&  (4\pi T)^{-{D\over 2}} \,
{\rm det}^{-\half}\Biggl\lbrack{\tan (eFT)\over eFT}\Biggr\rbrack
\qquad   (Spinor \,\, QED) \, .
\nonumber\\
\label{changedet}
\ear

\end{itemize}

\vspace{-6pt}
It should be remarked that the worldline formalism is closely related to the standard
Fock-Schwinger proper-time representation of propagators in
external fields \cite{fock,schwingervp}. 
Therefore the resulting integral representations have 
generally the same structure as the ones obtained by that method
(see, e.g., \cite{borska} and V. Skalozub's contribution to these proceedings). 
However, the worldline approach 
is more global in the sense that it applies directly to a whole loop, rather than
to the individual propagators making up the loop. This also implies that the
worldline integral representations can be written down
without fixing the ordering of the external legs along the loop. 
Another advantage is that
the use of the so-called ``Bern-Kosower substitution rule'' \cite{berkos} provides a simple
way of inferring the spinor loop integrands from the scalar loop ones. This
effectively circumvents the usual Dirac algebra manipulations.

In flat space, the gaussian integration of the worldline path integral can be done naively,
and no ill-defined expressions are produced. 
For applications to gravity, we need to generalize the path integrals
(\ref{Gammascal}), (\ref{grassmannpi}) to gravitational backgrounds.
Here we enter the realm of path integrals in curved spaces, a subject
notorious for its mathematical subtleties.

Naively, one would introduce background gravity by a simple
replacement of the kinetic term, 


\bear
S_0=\fourth\int_0^T d\tau  \dot x^2 &\to&\fourth \int_0^T d\tau  
\dot x^{\mu}g_{\mu\nu}(x(\tau))\dot x^{\nu}\, .
\label{putg}
\ear
After the usual linearization
$g_{\mu\nu} = \eta_{\mu\nu} + \kappa h_{\mu\nu}$
this would lead to a graviton vertex operator of the form


\bear
\varepsilon_{\mu\nu}\int_0^Td\tau \,\dot x^{\mu}
\dot x^{\nu}\,\e^{ik\cdot x} \, .
\label{vertopwrong}
\ear

\no
However, using this vertex operator in a naive gaussian path integration
immediately leads to ill-defined expressions  involving, e.g., 
$ \delta(0), \delta^2(\tau_i-\tau_j), \ldots 
\nonumber
$
A complete understanding of these difficulties, and of the steps which
have to be taken to solve them in the ``string-inspired'' framework,
has been reached only recently \cite{basvan,deb,bspvn,kleche,bacova}.
Here we can only briefly sketch the correct procedure for the spinless case; 
all the necessary
details and the generalization to spin half can be found in \cite{basvanbook}. 

\begin{enumerate}

\item
In curved space, the path integral measure is nontrivial. 
Exponentiate it as follows,


\bear
\fl\qquad
{\cal D} x &=& Dx \prod_{ 0 \leq \tau < T} \sqrt{\det g_{\mu\nu}(x(\tau))} 
=
Dx \int_{PBC} { D} a { D} b { D} c \;
{\rm e}^{- S_{gh}[x,a,b,c]}, \label{expmeasure}
\ear
with a ghost action 


\bear
S_{gh}[x,a,b,c]
= \int_{0}^{T} d\tau \; {1\over 4}g_{\mu\nu}(x)(a^\mu a^\nu 
+ b^\mu c^\nu) \, .\label{Sghost}
\ear
This modifies the naive graviton vertex operator (\ref{vertopwrong}) to


\bear
V_{\rm scal}^h[k,\varepsilon] = \varepsilon_{\mu\nu}\int_0^Td\tau \,
\Bigl[\dot x^{\mu}\dot x^{\nu}+a^{\mu}a^{\nu}+b^{\mu}c^{\nu}\Bigr]
\,\e^{ik\cdot x} \, .
\label{gravvertop}
\ear

\item
The correlators of these ghost fields just involve $\delta$ functions,


\bear
\langle a^{\mu}(\tau_1)a^{\nu}(\tau_2)\rangle
&=&
2\delta(\tau_1-\tau_2)\eta^{\mn}, \non\\
\langle b^{\mu}(\tau_1)c^{\nu}(\tau_2)\rangle
&=&
-4 \delta(\tau_1-\tau_2)\eta^{\mn} \, .
\label{wickghost}
\ear
The ghost field contributions will cancel all divergent or ill-defined terms. 

\item
This cancellation of infinities leaves finite ambiguities. 
From the point of view of one-dimensional quantum field theory, we are dealing
here with an UV divergent but super-renormalizable theory which requires only 
a small number of counterterms to remove all divergences.  
The coefficients of the counterterms
have to be fixed in a way which reproduces the known spacetime physics.

\item
As it turns out, these counterterms are regularization dependent,  and in general
noncovariant. This points to a violation of covariance by the regularization method.
Presently, the only known covariance-preserving regularization method is
one-dimensional dimensional regularization \cite{kleche,bacova}, which has only a single
covariant counterterm proportional to the curvature scalar  
($-R/4$ in the present notations).

\item
The zero mode fixing leads to further subtleties. The simplest possibility
would be to fix a point $x_0$ on the loop, $x(\tau)= x_0 + y(\tau)$. It leads to
the so-called DBC (``Dirichlet boundary conditions'') propagator for the
coordinate field which is known to yield the same effective Lagrangian
as would be obtained also by using the standard heat kernel expansion.   
For flat space calculations, the 
``string-inspired'' choice (\ref{zeromode}) is generally more convenient 
since it is the only one which leads to a worldline propagator
for the coordinate field depending only on $\tau_1-\tau_2$.  
It can be easily shown that the effective Lagrangians obtained in both
ways differ only by total derivative terms \cite{fhss}. 

This continues
to be true in curved space, however here those total derivative terms
turn out to be noncovariant in general, with only the DBC choice yielding
a manifestly covariant form of the Lagrangian. The noncovariance of the
total derivative terms present in the ``string-inspired'' approach poses no
problem in principle but in practice, since it invalidates the use of the
Riemann normal coordinate expansion, which is an almost indispensable
tool for this type of calculations. This remaining problem was solved
in \cite{bacozi}. There it was shown that, using Riemann normal coordinates
from the beginning and performing a BRST treatment of the symmetry corresponding to a 
shift of $x_0$, the difference between the two effective Lagrangians can be reduced
to manifestly covariant terms. This is achieved by the addition of further Fadeev-Popov type 
terms to the worldline Lagrangian in the ``string-inspired'' scheme.
Those terms are infinite in number but easy to determine order by order.

\end{enumerate}

\vspace{-6pt}

\no
All this can be generalized to the spin half case \cite{basvanbook}.
Thus a standard scheme of calculation is now
available for one-loop effective actions and amplitudes involving
scalar or spinor loop particles and background gravitational fields.
See \cite{dilmck} for some applications to effective actions and anomalies,
\cite{baszir,holsho} to graviton amplitudes.
More recently also worldline path integrals
representing vector and antisymmetric tensor particles coupled to background
gravity have been constructed \cite{babegi}.

\vspace{-5pt}

\vfill\eject

\section{Photon-graviton polarization tensor in a constant field}


Returning to the one-loop photon-graviton amplitude
in a constant electromagnetic field $F_{\mu\nu}$, we will now sketch
its calculation for the scalar loop case. According to the above, this amplitude 
can be represented by the following expression,


\bear
\fl\quad
\varepsilon_{\mu\nu}\Pi^{\mu\nu ,\alpha}_{\rm scal}(k)\varepsilon_{\alpha} &=&
{ie\kappa\over 4}
\TintmD
{\rm det}^{-{1\over 2}}
\Biggl[{{\rm sin}({\cal Z})\over {\cal Z}}\Biggr]
\Bigl\langle
V^h_{\rm scal}[k,\varepsilon_{\mu\nu}] V^A_{\rm scal}[-k,\varepsilon_{\alpha}]
\Bigr\rangle
\nonumber\\
\label{hAwlwick} 
\ear
where ${\cal Z}_{\mn} \equiv eTF_{\mn}$ and 
$V^{A,h}_{\rm scal}$
are the photon and graviton vertex operators 
(\ref{photvertop}), (\ref{gravvertop}).
Using the Wick contraction rules (\ref{GB}), (\ref{wickghost})
yields ($\overline{\cal G}_{B12}:= {\cal G}_{B12}-{\cal G}_{B11}$ etc.)

\bear
\fl\qquad
\Pi^{\mu\nu ,\alpha}_{\rm scal}(k)\!\!\! &=&
{e\kappa\over 4 (4\pi)^{D\over 2}}
\int_0^{\infty} {dT\over T^{1+{D/2}}}\,\e^{-m^2T}
{\rm det}^{-{1\over 2}}
\Biggl[{{\rm sin}({\cal Z})\over {\cal Z}}\Biggr]
\int_0^Td\tau_1\int_0^Td\tau_2
\,\e^{-k\cdot \overline {\cal G}_{B12}\cdot k}
I^{\mu\nu ,\alpha}_{\rm scal},
\nonumber\\
\fl\qquad
I^{\mu\nu ,\alpha}_{\rm scal} \!\!\!
&=&
-\Bigl({\ddot {\cal G}}_{B11}^{\mn}-2\delta_{11}\eta^{\mn}\Bigr)
\Bigl(k\cdot \,{\overline {\dot {\cal G}}}_{B12}\Bigr)^{\alpha}
-\Bigl[ {\ddot {\cal G}}_{B12}^{\mu\alpha}
\Bigl(\,{\overline {\dot {\cal G}}}_{B12}\cdot k\Bigr)^{\nu}
+ \exmn \Bigr] \non\\&&
+ \Bigl(\,{\overline {\dot {\cal G}}}_{B12}\cdot k\Bigr)^{\mu}
\Bigl(\,{\overline {\dot {\cal G}}}_{B12}\cdot k\Bigr)^{\nu}
\Bigl(k\cdot \,{\overline {\dot {\cal G}}}_{B12}\Bigr)^{\alpha}.
\nonumber\\
\label{wickresult}
\ear


\no
The $T$ integral has an UV divergence at the lower limit.
Using dimensional regularization, this divergence takes the form


\bear
\Pi_{\rm scal, div}^{\mn,\alpha}(k)
&=&
{ie^2\kappa\over 3(4\pi)^2}{1\over D-4}C^{\mn,\alpha}
\label{UVdiv}
\ear
\bigskip

\no
where $C^{\mn,\alpha}$ is the tree level vertex (\ref{defC}). Adding the corresponding
counterterm yields the
renormalized vacuum polarization tensor $\bar\Pi^{\mu\nu ,\alpha}_{\rm scal}$ 
obeying the usual renormalization condition $\bar\Pi^{\mu\nu ,\alpha}_{\rm scal}(k=0)=0$. 

Next, appropriate photon and graviton polarizations have to be selected, where it turns
out to be convenient to use the photon polarization vectors also to construct the
graviton polarization tensors:

\bigskip

\no
{\it Photon:}   $\quad\varepsilon_{\perp}, \varepsilon_{\parallel}$, 

\medskip\no
{\it Graviton:} $
\varepsilon^{\oplus\mu\nu} = \varepsilon^{\perp\mu}\varepsilon^{\perp\nu}
- \varepsilon^{\parallel\mu}\varepsilon^{\parallel\nu}, \quad
\varepsilon^{{\otimes}\,\mu\nu} = \varepsilon^{\perp\mu}\varepsilon^{\parallel\nu}
+ \varepsilon^{\parallel\mu}\varepsilon^{\perp\nu}. 
$

\bigskip

\no
Here we have assumed a Lorentz system such that $\bf B$ and $\bf E$ are collinear,
and the subscripts on the photon polarization vectors refer to the same direction.
Further, no information is lost by assuming that the photon propagation is perpendicular
to the field \cite{pg2}.

\vfill\eject\noindent
With these conventions, the components of the tree level amplitude become

\bear
C^{\oplus\perp}
&=&
-2B\omega \, ,\nonumber\\
C^{\oplus\parallel}
&=&
2E \omega\, , \nonumber\\
C^{\otimes\perp}
&=&
-2 E \omega  \, ,\nonumber\\
C^{\otimes\parallel} 
&=& -2B \omega \, .\nonumber\\
\label{poldecomptree}
\ear
\vspace{-25pt}

\no
Here $\omega = k^0=\vert {\vec k} \vert$ denotes the
photon/graviton energy.
Finally, it is convenient to normalize the loop amplitude by the tree level one,
making the amplitude dimensionless:


\bear
\hat \Pi^{Aa}_{\rm scal}
(\hat\omega,\hat B,\hat E) &\equiv& 
{\rm Re}\Biggl({\bar \Pi^{Aa}_{\rm scal}(\hat\omega,\hat B,\hat E)
\over -{i\over 2}\kappa C^{Aa}}\Biggr)
\non\\
\label{normamp}
\ear
\no
($A=\oplus,\otimes$, $a= \perp, \parallel$).
Here we have further introduced the
dimensionless variables
$\hat\omega = {\omega\over m}$ ,
$\hat B = {eB\over m^2}$ , 
$\hat E = {eE\over m^2}$ .

The spinor loop calculation proceeds completely analogously, 
just with some additional terms coming from the evaluation of the
spin path integral (\ref{grassmannpi}).

At this stage, the four independent 
components of the scalar or spinor loop amplitude are given in terms of two-parameter
integrals, with integrands involving trigonometric functions of the
proper times and external parameters. Let us write down here these integrals for
the case of a spinor loop and a purely magnetic field: 

\bear
\hat \Pi^{Aa}_{\rm spin}
(\hat\omega,\hat B)
&=& \alpha \,{\rm Re} \int_0^{\infty}{d\hat s\over \hat s}\,\e^{-i\hat s}\int_0^1dv\,
\hat \pi^{Aa}_{\rm spin}(\hat s,v,\hat\omega,\hat B)
\non\\
\label{Pihatspin}
\ear

\vspace{-10pt}

\bear
\fl\quad
\hat \pi_{\rm spin}^{\oplus\perp}
=
-{1\over 4\pi}\biggl\lbrace{z\over\tanh(z)}
\exp\Bigl[z\Bigl({\bar A_{B12}\over z}+\half (1-v^2)\Bigr){\hat\omega^2\over 2\hat B}\Bigr]
\non\\
\fl\qquad\qquad\times
\Bigl[
(S_{B12})^2 - (S_{F12})^2 + (A_{F12})^2 
-\Bigl(\bar A_{B12}+A_{F11}\Bigr)\Bigl(\bar A_{B12}+{1\over z}+A_{F11}\Bigr)
\non\\
\fl\qquad\qquad
+\bar A_{B12}\Bigl((S_{B12})^2 - (S_{F12})^2 -(\bar A_{B12}+A_{F11})^2
+(A_{F12})^2-vS_{B12} +S_{F12}\Bigr)
{\hat\omega^2\over 2\hat B}\Bigr]
+{4\over 3}
\biggr\rbrace \, ,
\non\\
\fl\quad
\hat \pi_{\rm spin}^{\otimes\parallel}
=
-{1\over 4\pi}\biggl\lbrace{z\over\tanh(z)}
\exp\Bigl[z\Bigl({\bar A_{B12}\over z}+\half (1-v^2)\Bigr){\hat\omega^2\over 2\hat B}\Bigr]
\non\\
\fl\qquad\qquad\quad\times
\biggl[vS_{B12} - S_{F12} -{1\over z}\Bigl(\bar A_{B12}+A_{F11}\Bigr)
+\bar A_{B12}\Bigl(vS_{B12}-S_{F12}+1-v^2\Bigr){\hat\omega^2\over 2\hat B}\biggr]
+{4\over 3}
\biggr\rbrace \, ,
\non\\
\fl\quad
\hat \pi_{\rm spin}^{\oplus\parallel}
= 0 \, , \non\\
\fl\quad \hat \pi_{\rm spin}^{\otimes\perp} = 0 \, .\non\\
\label{hatpimagspin}
\ear 
Here $\hat s  = -im^2 T$, $z=i\hat B \hat s$, and the integrand involves 
the standard worldline coefficient functions \cite{vv} 

\bear
S_{B12} &=&
{\sinh(zv)\over \sinh(z)} ,
\non\\
A_{B12} &=&
{\cosh(z v)\over 
\sinh(z)}-{1\over z} ,\nonumber\\
A_{B11} &=& A_{B22} =
\coth(z)-{1\over z}, \non\\ 
\bar A_{B12}&=& A_{B12}-A_{B11}
= {\cosh(z v)-\cosh(z)\over 
\sinh(z)},\non\\
S_{F12} &=& 
{\cosh(zv)\over \cosh(z)} ,
\non\\
A_{F12} &=& 
{\sinh(z v)\over 
\cosh(z)}  ,\non\\
A_{F11} &=& A_{F22} = {\tanh}(z) \, .\non\\
\label{defSA}
\ear
The parameter $v$ is related to the original proper-time variables $\tau_{1,2}$ by 
 $v=1-2\tau_1/T$ (the translation invariance of the worldline correlators has been
 used to set $\tau_2=0$).
 
\no
See \cite{pg2} for the scalar loop and general constant field cases.

\vspace{-8pt}

\section{Properties, special cases}

\vspace{-5pt}

\no
Let us now discuss some properties and limiting cases of the amplitude:


\no
\underline{\it Ward identities:}
The gauge Ward identity for this amplitude gives the familiar transversality in the
photon index,


\bear
k_{\alpha}\Pi^{\mn,\alpha}(k) &=& 0\label{wardgauge} \, .
\ear
The gravitational Ward identity, derived from invariance under infinitesimal
reparametrizations, connects $\Pi^{\mn,\alpha}$ with 
the corresponding photon-photon polarization tensor $\Pi^{\mu,\alpha}(k)$, 

\vspace{-18pt}

\bear
k_{\mu}\Pi^{\mn,\alpha}(k) &=& {i\over 2}\kappa F^\nu{}_\mu\Pi^{\mu,\alpha}(k) \, .
\label{wardgrav}
\ear
(Similarly, non-transversality was recently found for the gluon polarization tensor in
a chromomagnetic background field \cite{bogrsk}.)

\medskip

\no
\underline{\it Selection rules:} CP invariance implies the following selection rules 
for the photon-graviton conversion amplitudes \cite{rafsto}:

\begin{itemize}

\item
For a  purely magnetic field  $\varepsilon^{\oplus}$  couples only
to  $\varepsilon^{\perp}$   and  
 $\varepsilon^{\otimes}$  only
to  $\varepsilon^{\parallel}$.

\item
For a  purely electric field  $\varepsilon^{\oplus}$  couples only
to  $\varepsilon^{\parallel}$  and  
 $\varepsilon^{\otimes}$  only
to  $\varepsilon^{\perp}$.

\end{itemize}

\no
This is borne out by the explicit calculation (see (\ref{poldecomptree}), (\ref{hatpimagspin})).

\bigskip

\no
\underline{\it Pair creation thresholds:} 
In the purely magnetic case the amplitudes are real
for small $\omega$, since the magnetic field is not capable
of pair production. The pair creation thresholds $\omega_{\rm cr}$ turn out
to be identical with the ones for the corresponding 
photon-photon cases:


\bear
\hat\omega_{\rm cr,scal}^{\oplus \perp} =
\hat\omega_{\rm cr,scal}^{\otimes \parallel}  &=& 2\sqrt{1+\hat B} \, ,
\non\\
\hat\omega_{\rm cr,spin}^{\oplus\perp} &=& 1+\sqrt{1+2\hat B} \, ,
\label{thresholds}\\
\hat\omega_{\rm cr,spin}^{\otimes\parallel} &=& 2 \, . \nonumber
\ear

\bigskip

\no
\underline{\it Calculable cases:} The magnetic case is also much more
amenable to an explicit calculation of the parameter integrals. 
In \cite{pg2} we have given a detailed analysis of the following
regions in parameter space (with ${\bf E}=0$):

\begin{itemize}

\item
For photon/graviton energies below threshold the parameter integrals
are suitable for a straightforward numerical evaluation.

\item
For arbitrary  $\omega$ but small $B$ the two-parameter integrals
can be reduced to one-parameter integrals over Airy functions.

\item
For $\omega < \omega_{\rm cr}$ and large $B$  one finds the
asymptotic behaviour


\bear
\hat \Pi^{Aa}_{\rm scal}
(\hat\omega,\hat B) &\stackrel{\hat B\to\infty}{\sim}&
-{\alpha\over 12\pi}\ln (\hat B) \, ,
 \non\\ 
\hat \Pi^{Aa}_{\rm spin} 
(\hat\omega,\hat B) &\stackrel{\hat B\to\infty}{\sim}& 
-{\alpha\over 3\pi}\ln (\hat B) \, . \nonumber\\
\label{asymp}
\ear


\no
These leading asymptotic terms can be directly related to the
corresponding UV counterterms, which is another property
known from the photon-photon case \cite{ritus}. 

\item
In the zero energy limit, the amplitudes relate to the magnetic
Euler-Heisenberg Lagrangians ${\cal L}_{\rm scal,spin}^{\rm EH}(B)$:


\bear
\hat\Pi_{\rm scal,spin}^{\oplus \perp}(\hat\omega=0,\hat B) 
&=& 
-{2\pi\alpha\over m^4} \Bigl({1\over \hat B}{\pa\over\pa \hat B} + {\pa^2\over \pa \hat B^2}\Bigr)
\,{\cal L}_{\rm scal,spin}^{\rm EH}(\hat B) \, ,
\non\\
\hat\Pi_{\rm scal,spin}^{\otimes \parallel}(\hat\omega=0,\hat B)
&=& 
-{4\pi \alpha\over m^4}
{1\over \hat B}{\pa\over\pa \hat B}\, {\cal L}_{\rm scal,spin}^{\rm EH}(\hat B) \, .
\nonumber\\
\label{EHtoamp}
\ear


\no
The identities (\ref{EHtoamp}) have also been derived by  
Gies and Shaisultanov using a different approach \cite{giesha}.

\end{itemize}

\vspace{-12pt}

\section{Conclusions}

\vspace{-6pt}

The calculation presented here is the 
first calculation of the photon-graviton vacuum polarization
in a constant electromagnetic field, and also the 
first state-of-the-art application of the ``string-inspired'' worldline formalism to an
amplitude involving gravitons. Although it was not possible here
to go into detail, it should be emphasized that
in this formalism this calculation is only
moderately more difficult than the photon-photon polarization
in the field. Moreover, we have also shown that the properties of
the photon-graviton polarization tensor are very similar to the ones of the
photon-photon one. 
We expect that even the graviton-graviton
case will be quite feasible in this formalism. 
In a future sequel, we intend to analyze this case at the
same level of the photon-graviton one, and to study the complete
one-loop photon-graviton dispersion relations (\ref{disploop}).




\end{document}